\documentclass[conference]{IEEEtran}
\IEEEoverridecommandlockouts
\usepackage{cite}
\usepackage{amsmath,amssymb,amsfonts}
\usepackage{algorithmic}
\usepackage{graphicx}
\usepackage{textcomp}
\usepackage{xcolor}
\usepackage{url}
\def\BibTeX{{\rm B\kern-.05em{\sc i\kern-.025em b}\kern-.08em
    T\kern-.1667em\lower.7ex\hbox{E}\kern-.125emX}}
\begin{document}

\title{Theoretical Analysis on Block Time Distributions in Byzantine Fault--Tolerant Consensus Blockchains
\thanks{In the course of this research, the author had fruitful comments from Prof. Tomoyuki Shirai of Kyushu Univ., Prof. Hiroyoshi Miwa of Kwansei Gakuin Univ., and Mr. Hongru He of Chiba Institute of Technology. This research was partially supported by the Grant-in-Aid for Scientific Research (C) (No. 23K11086) from the Japan Society for the Promotion of Science.}
}

\author{\IEEEauthorblockN{Akihiro Fujihara}
\IEEEauthorblockA{\textit{Graduate School of Engineering, Chiba Institute of Technology} \\
2-17-1 Tsudanuma, Narashino, Chiba 275--0016, Japan \\
akihiro.fujihara@p.chibakoudai.jp}
}

\maketitle

\begin{abstract}
Some blockchain networks employ a distributed consensus algorithm featuring Byzantine fault tolerance. 
Notably, certain public chains, such as Cosmos and Tezos, which operate on a proof--of--stake mechanism, have adopted this algorithm. 
While it is commonly assumed that these blockchains maintain a nearly constant block creation time, empirical analysis reveals fluctuations in this interval; this phenomenon has received limited attention. 
In this paper, we propose a mathematical model to account for the processes of block propagation and validation within Byzantine fault--tolerant consensus blockchains, aiming to theoretically analyze the probability distribution of block time. 
First, we propose stochastic processes governing the broadcasting communications among validator nodes. 
Consequently, we theoretically demonstrate that the probability distribution of broadcast time among validator nodes adheres to the Gumbel distribution. 
This finding indicates that the distribution of block time typically arises from convolving multiple Gumbel distributions. 
Additionally, we derive an approximate formula for the block time distribution suitable for data analysis purposes. 
By fitting this approximation to real--world block time data, we demonstrate the consistent estimation of block time distribution parameters. 
\end{abstract}


\section{Introduction}\label{sec:introduction}

Blockchains typically generate blocks containing multiple transactions, which are then shared among validator nodes through broadcast communications. 
This process ensures the correctness of the block before it is permanently added to the blockchain. 
The time taken for this entire sequence is known as the block time. 
Block time comprises three main stages: block creation time, block propagation time through broadcast communications to other validator nodes, and block verification time by validator nodes. 
A shorter block time leads to faster transaction approval, enhancing the convenience of the blockchain. 
However, achieving consensus among geographically dispersed nodes regarding which blocks to include in the blockchain is crucial. 
Setting the block time too short may jeopardize consensus and lead to blockchain forks, rendering the system unstable. 
Therefore, block times are typically adjusted empirically to strike a balance between convenience and security. 

This study primarily focused on Byzantine fault--tolerant consensus blockchains such as Cosmos\cite{cosmos}, Tezos\cite{tezos}, and various others\cite{SNG2020}. 
Block times are typically determined empirically for each blockchain system. 
For instance, Cosmos has block times of 6--7s, while Tezos operates with approximately 15s block times. 
These blockchain systems utilize distributed consensus algorithms -- Tendermint\cite{BKM2018} (or CometBFT\cite{IDA}) for Cosmos and Tenderbake\cite{ACPRTZ2021} for Tezos. 

The distributed consensus algorithm with Byzantine fault tolerance establishes consensus through a communication protocol grounded in Practical Byzantine Fault Tolerance (PBFT)\cite{CL1999}. 
PBFT seeks to realize consensus via three stages of broadcast communications: PROPOSE, PRE-COMMIT, and COMMIT. 
Validator nodes, a specific set of nodes involved in creating and approving a block, vote on whether to approve the block for inclusion in the blockchain. 
Once consensus is reached, there is no possibility of reversal, thereby finalizing transactions within the block as soon as it is incorporated into the blockchain. 
Consequently, Byzantine fault--tolerant blockchains typically do not experience forks. 

Byzantine fault--tolerant consensus blockchains typically adopt proof of stake (PoS)\cite{pos}, wherein only nodes that have staked a certain amount of cryptoassets can engage in block creation and validation. 
Specifically, Cosmos sets the number of validator nodes at $N(=3f+1)=175$\cite{IDA}. 
This restricts participation to a select group of validator nodes, comprising the top 175 holders of staked cryptoasset, who contribute to the block creation and validation processes. 
Cosmos leverages staked cryptoassets to facilitate secure consensus formation, provided that no more than $f=58$ validator nodes collude and mount simultaneous attacks. 
Given that PBFT is a distributed consensus algorithm utilized in consortium chains, while Cosmos and Tezos operate as public chains, their consensus mechanisms closely resemble those of consortium chains.

Ethereum 2.0\cite{CPNX2020} also adopts a PoS--based consensus, yet it diverges from Byzantine fault--tolerant consensus blockchains. 
Employing a distributed consensus algorithm that melds GHOST and Casper\cite{BHKPQRSWZ2000}, Ethereum 2.0 does not achieve immediate transaction finalization upon block inclusion, potentially leading to blockchain forks. 
This discrepancy stems from differing project emphases: Ethereum 2.0 strives for a large validator node count ($N \simeq 488,000$)\cite{GSR2022} to amplify decentralization, whereas Cosmos opts for a smaller validator node count ($N=175$)\cite{IDA}. 
Ethereum 2.0 prioritizes augmenting validator nodes to bolster decentralization beyond Byzantine fault--tolerant consensus blockchains, albeit at the expense of blockchain security, allowing for potential forks.
Cosmos, conversely, prioritizes cross-chain interoperability, facilitating token exchange and secure data transfer across multiple blockchains. 
Thus, Cosmos aims to enhance security, enabling swift block finalization without the occurrence of blockchain forks, albeit at the expense of decentralization through a reduction in validator node count. 
These disparities within the same PoS--based blockchain serve as an interesting illustration of the trade-off between security and decentralization\cite{AHMKS2020}. 
Notably, Ethereum 2.0 also implements a mechanism for block finalization spanning 2 epochs (= 64 slots = 12.8 minutes), involving voting among validator nodes, similar to the consensus algorithms of Byzantine fault--tolerant consensus blockchains\cite{BHKPQRSWZ2000}.

\begin{figure}[t!]
  \centering
  \includegraphics[width=0.50\textwidth]{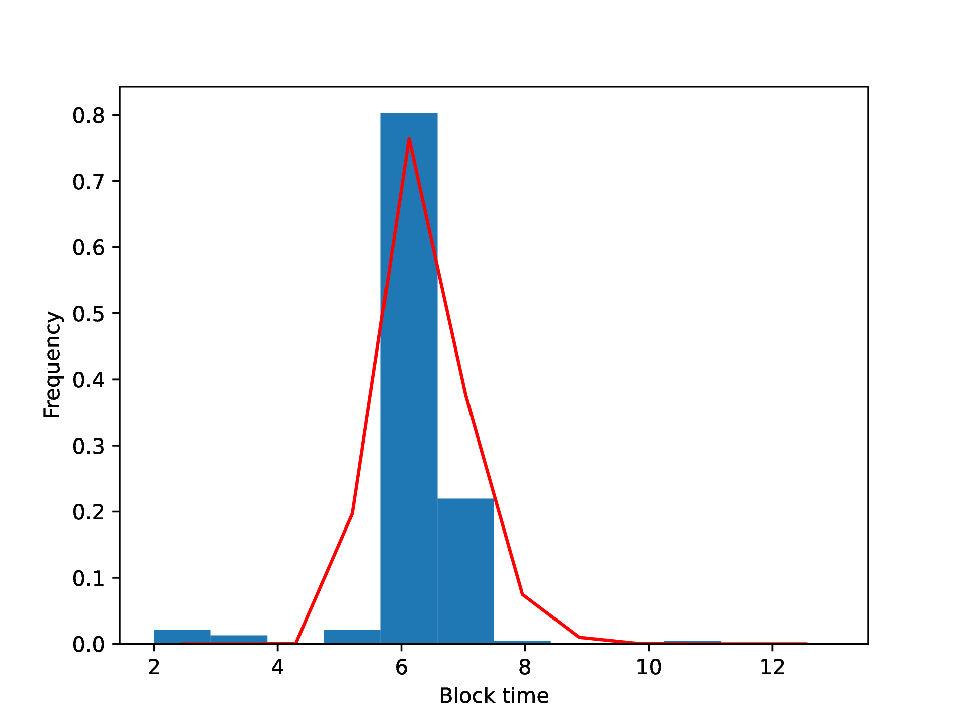}
  \includegraphics[width=0.50\textwidth]{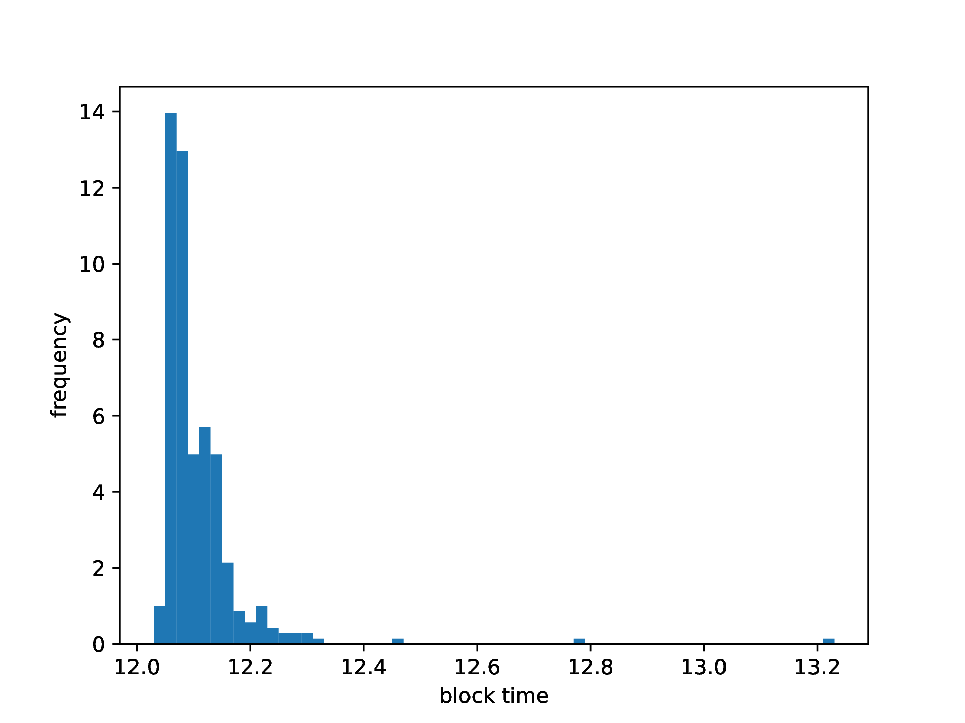}
  \caption{Histograms of block time in Byzantine fault--tolerant consensus blockchains and Ethereum 2.0. 
	Those for Cosmos and Ethereum 2.0 are depicted at the top and bottom of the figure, respectively. 
	The solid red line at the top of the figure is provided by non-linear fitting using a theoretical result in Eq. (\ref{eq:approximate_result3}) to be explained later.}
  \label{fig:data_blocktime}
\end{figure}

We collected actual data on block times in PoS--based blockchains, which are depicted in histograms presented in Fig.~\ref{fig:data_blocktime}.
Block time data for Cosmos was collected from a blockchain explorer (\url{https://cosmos.explorers.guru/}), and data for Ethereum 2.0 were sourced from Etherscan \cite{etherscan}. 
As shown in Fig.~\ref{fig:data_blocktime}, sharp peaks are evident around typical block times in the histograms (6--7s in Cosmos, and approximately 12s in Ethereum 2.0). 
However, Fig.~\ref{fig:data_blocktime} also indicates minor fluctuations around these typical block times.
These fluctuations are presumed to stem from the inherent nature of the distributed consensus algorithm, although they lack a comprehensive theoretical understanding. 
If the relationship between the number of validator nodes and metrics on the block time is mathematically clarified, the average block time that has been decided empirically observing the balance between the transaction processing efficiency and the consistency of the blockchain can be determined by theoretical grounds without performing real--world experiments, which is a motivation of this research. 

In contrast, a prior study reported that block time in Cosmos adheres to the Erlang distribution\cite{WHLWL2023}. 
Nonetheless, the study was based on the assumption that the verification time required for transaction incorporation into a block in PoS--based blockchains, which do not rely on proof of work (PoW), follows an exponential distribution. 
However, there exists no detailed explanation of a theory that sufficiently supports this assumption. 
While the exponential distribution entails a probability distribution where the mean and standard deviation coincide, it remains unconfirmed whether the mean and standard deviation, equivalent to the magnitude of fluctuation, align. 

In this paper, we propose a mathematical model to theoretically elucidate the time required for block propagation and verification in Byzantine fault--tolerant consensus blockchains. 
Through theoretical analysis of this model, we derive analytical results capable of elucidating block time fluctuations in Cosmos. 
The primary contributions of this study are as follows: 
\begin{itemize}
	\item To theoretically analyze the time required for block propagation and verification, we propose a stochastic model that describes the process of broadcasting blocks among validator nodes. 
		Through theoretical analysis of this model, we demonstrate that the broadcast time of blocks follows the Gumbel distribution\cite{gumbel1954} when the number of validator nodes is sufficiently large. 
	\item We demonstrate that the verification time (or attestation time) of blocks follows a probability distribution obtained by convolving two independent Gumbel distributions. 
		Furthermore, we establish that the block time distribution follows a probability distribution obtained by convolving three independent Gumbel distributions. 
	\item We derive an approximate representation of the block time distribution, offering a valuable tool for theoretical examination of block time through fitting with actual block time data. 
	\item By applying the derived approximate representation to real block time data from Cosmos, we obtain analytical results that quantitatively account for the fluctuation of block time observed in the histogram. 
\end{itemize}

The structure of this paper is organized as follows. 
In Section \ref{sec:relatedworks}, we present existing works relevant to this research. 
Section \ref{sec:theory} introduces a stochastic model that delineates the process of broadcasting blocks among validator nodes. 
Through theoretical analysis, we derive mathematical formulas concerning the probability distribution of broadcast time, validation time, and block time from this stochastic model. 
In Section \ref{sec:discussion}, we analyze real block time data by fitting them with the results of our theoretical analysis to estimate parameters in the block time distribution. 
Finally, Section \ref{sec:conclusion} summarizes the findings of this study and outlines potential future work.

\section{Related works}\label{sec:relatedworks}

The fluctuation of block time in the Nakamoto consensus, as observed in the case of Bitcoin, can be theoretically analyzed using queuing theory\cite{KK2019}. 
It is commonly understood that block time follows an exponential distribution when PoW is employed for the consensus algorithm\cite{YF2021a,YF2021b,Fujihara2023a,Fujihara2023b,Fujihara2020,Fujihara2020j}. 
Nakamoto consensus does not finalize blocks because it is a stochastic nature, supporting only eventual consistency. 
However, in the case of Bitcoin, there's a convention known as the Six-confirmation rule, assuming pseudo-finalization. 
According to this rule, the probability of six consecutive new blocks being created within a single block time is considered sufficiently small, practically finalizing transactions. 

Since the fluctuation of the block time in Byzantine fault--tolerant consensus blockchains is negligibly small, there are few papers which mathematically discuss the probability distribution of the block time. 
We searched Google scholar with terms such as ``byzantine fault tolerant consensus blockchain block time distribution'' and ``PBFT consensus latency distribution'' and found few research papers discussing probability distributions of block times mathematically. 
Probably, for most researchers, there was no inconvenience in considering the block time as a constant, it may have been out of interest other than the following few:  
A prior study has reported that block times in Cosmos adhere to the Erlang distribution\cite{WHLWL2023}. 
However, this study was based on the assumption that the time it takes for a transaction to be included in a block follows the exponential distribution which is reasonable only in the case of PoW--based blockchains, lacking sufficient theoretical support. 

Cosmos, a blockchain ecosystem fostering cross-chain interoperability, facilitates message relay across multiple chains. 
A study has indicated that the time taken for message relay via Cosmos's Inter-Blockchain Communication (IBC) conforms to a log-normal distribution, ranging from approximately 28--139s with an average IBC message relay time of approximately 55s \cite{EKJ2023}. 
This research collected data on IBC message relays between Cosmos Hub and Osmosis to visualize their time distribution. 
Note that the IBC message relay time consists of three block--time segments: Generation--to--Transfer, Transfer--to--Receive, and Receive--to--Acknowledgment interval times. 
Despite conducting statistical tests such as the Kolmogorov--Smirnov and the Shapiro--Wilk tests, null hypothesis of a normal distribution was not rejected. 
Therefore, it remains uncertain whether the distribution of IBC message relay time is truly log-normal. 
Furthermore, the theoretical examination as to why it follows a log-normal distribution has not been explored using mathematical models.

\section{Theoretical analysis}\label{sec:theory}

\subsection{Components of block time}

We assume that block time can be decomposed into the sum of the following three components: 
\begin{equation}
	\tau_{block} = \tau_{create} + \tau_{broadcast} + \tau_{attest}, 
\end{equation}
where $\tau_{create}$ denotes the time it takes for a validator node to first create a candidate block, $\tau_{broadcast}$ denotes the time it takes to broadcast the candidate block to other validator nodes, and $\tau_{attest}$ denotes the time it takes for validator nodes to verify the received candidate blocks independently. 

\begin{figure}[t!]
  \centering
  \includegraphics[width=0.50\textwidth]{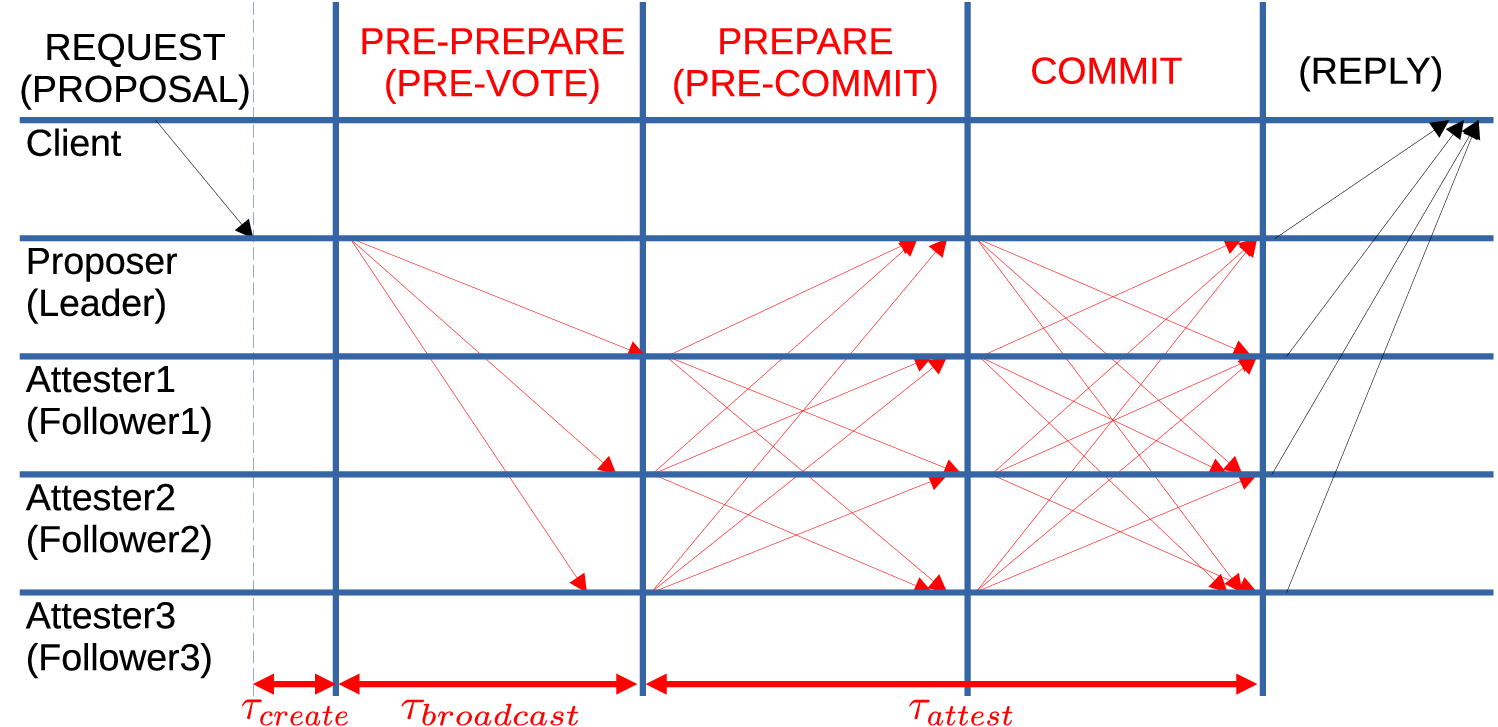}
  \caption{Three phases of repeating broadcasts of block information among validator nodes in Byzantine fault--tolerant consensus blockchains}
  \label{fig:bft_consensus}
\end{figure}

As the Nakamoto consensus employs PoW, the bottleneck time is the duration required to create a candidate block, \textit{i.e.}, 
\begin{equation}
	\tau_{create} \gg \tau_{broadcast} + \tau_{attest}.
\end{equation}
Given that the time taken by PoW follows an exponential distribution, the block time likewise follows this distribution\cite{KK2019}. 
In contrast, in Byzantine fault--tolerant consensus blockchains, the bottleneck time is not the duration to create a candidate block, as there is no PoW process involved in creating candidate blocks. 
Instead, the bottleneck time emerges in verifying candidate blocks through repeated broadcasts among numerous validator nodes. 
As illustrated in Fig.~\ref{fig:bft_consensus}, the distributed consensus algorithm with Byzantine fault tolerance entails three phases of block information broadcast repetition. 
When the number of validator nodes is $N$, the complexity of communication volume for these phases escalates to $O(N^2)$.
These phases are regarded as the bottleneck time, \textit{i.e.}, 
\begin{equation}
	\tau_{broadcast} + \tau_{attest} \gg \tau_{create}. 
\end{equation}
These relationship indicates a reversal in the bottleneck time between PoW--based and Byzantine fault--tolerant consensus blockchains. 
In essence, constructing a mathematical model capable of addressing fluctuations in the time taken for broadcast communications between validator nodes and analyzing it theoretically would enable us to comprehend the block time distribution.

\subsection{Stochastic processes on broadcast communications of candidate block between validator nodes}

First, we present a stochastic model delineating the time needed to broadcast a candidate block from a Proposer (Leader) to other Attesters (Followers), akin to the PRE-PREPARE (PRE-VOTE) phase depicted in Fig.~\ref{fig:bft_consensus}. 
We consider a Markov process with $N$ states for the number of validator nodes ($\mathbf{S} = \{ 1, 2, \cdots, N \}$). 
The initial state $1$ signifies a state where one validator has created a candidate block before the broadcast commences. 
Each state $(1 <) i (\le N)$ represents a state where $i$ validator nodes holding the candidate block broadcast it from the validator that initially created the block. 

\begin{figure}[t!]
  \centering
  \includegraphics[width=0.50\textwidth]{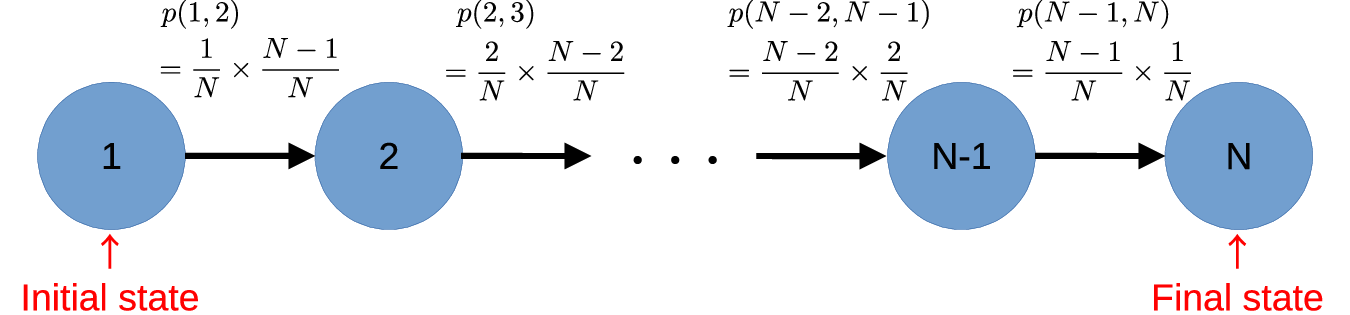}
  \caption{State transition diagram in the process on broadcast communications of candidate block between validator nodes}
  \label{fig:states_transitions}
\end{figure}

The state transition diagram of the processes is depicted in Fig.~\ref{fig:states_transitions}.
The transition probabilities between these states are defined by 
\begin{equation}
        p(i, i+1) = \frac{i}{N} \times \frac{N-i}{N} \hspace{5mm} (i=1, 2, \cdots, N-1). \label{eq:transition_probability}
\end{equation}
In Eq.~(\ref{eq:transition_probability}), the term $i/N$ represents the density of validator nodes that have already received the candidate block, while $(N-i)/N$ represents the density of validator nodes yet to receive it. 
The transfer speed is expressed as the product of these factors. 
Within these transition probabilities, we assume that at each time step, validator nodes that have received the candidate block transmit it to randomly selected other validator nodes, with this transfer process executed in parallel if multiple validator nodes hold the block. 

Implicit in this analysis is the assumption of a fully connected peer--to--peer network among validator nodes. 
In Byzantine fault--tolerant consensus blockchains, the number of validator nodes, $N$, is typically limited to a few hundred owing to the use of PBFT--based consensus algorithms. 
For instance, in Cosmos, as afore mentioned, $N=175$. 
Additionally, to expedite consensus formation, it is imperative to establish direct connections between validator nodes to maximize block transfer rates. 
Given these considerations, it is reasonable to assume that the network topology among validator nodes closely interconnected. 

Utilizing this stochastic model, we obtain the broadcast time $T$ as 
\begin{equation}
        T = \sum_{i=1}^{N-1} T_i, 
\end{equation}
where $T_i$ is a random variable representing the sojourn time of state $i$. 
As the state transition described in Eq.~(\ref{eq:transition_probability}) is assumed for each time step, the sojourn time of each state $(1 \le) i (\le N-1)$ is typically distributed according to the geometric distribution\cite{DKLM2005} by a simple Markov process. 
\begin{equation}
        f(j; i) = (1 - p(i, i+1) )^{j-1} p(i, i+1) \hspace{5mm} (j \ge 1). 
\end{equation}
The average of $T_i$ is calculated as follows: 
\begin{equation}
        E[T_i] = \frac{1}{p(i, i+1)} = \frac{N^2}{i(N-i)} = N \left( \frac{1}{i} + \frac{1}{N-i} \right).
\end{equation}
Therefore, the average broadcast time, $E[T]$, can be calculated as follows: 
\begin{eqnarray}
        E[T] &=& E\left[ \sum_{i=1}^{N-1} T_i \right] = \sum_{i=1}^{N-1} E[T_i] = N \sum_{i=1}^{N-1} \left( \frac{1}{i} + \frac{1}{N-i} \right) \nonumber \\
             &=& 2N \sum_{i=1}^{N-1} \frac{1}{i} = 2N \times H(N) = 2N \ln N + O(N), \label{eq:T_average}
\end{eqnarray}
where $H(N) = \ln N + O(1)$. 
This analysis indicates that when the number of validator nodes is $N$, the average broadcast time is determined on the order of $N \ln N$. 

Furthermore, the variance of broadcast time, $Var[T_i]$, can be calculated as follows: 
\begin{eqnarray}
        Var[T_i] &=& \frac{1 - p(i, i+1)}{p(i, i+1)^2} = \frac{1}{p(i, i+1)^2} - \frac{1}{p(i, i+1)} \nonumber \\
                 &=& N^2 \left( \frac{1}{i} \right)^2 + N^2 \left( \frac{1}{N-i} \right)^2 \nonumber \\
		 & & + N  \left( \frac{1}{i} + \frac{1}{N-i} \right). 
\end{eqnarray}
Therefore, the variance $Var[T]$ can be calculated as follows under the assumption that $T_i$ are independent of each other. 
\begin{eqnarray}
        Var[T] &=& Var\left[ \sum_{i=1}^{N-1} T_i \right] = \sum_{i=1}^{N-1} Var[T_i] \nonumber \\
               &=& N^2 \sum_{i=1}^{N-1} \left\{ \left( \frac{1}{i} \right)^2 + \left( \frac{1}{N-i} \right)^2  \right\} \nonumber \\
               & & + N \sum_{i=1}^{N-1} \left( \frac{1}{i} + \frac{1}{N-i} \right) \nonumber \\
               &\le& 2 \times \frac{\pi^2 N^2}{6} + 2N \times H(N) \nonumber \\
	       &=& \frac{\pi^2}{3} N^2 + O(N \ln N). \label{eq:T_variance}
\end{eqnarray}
Therefore, the standard deviation of $T$ is found to be $O(N)$.
\begin{equation}
        \sigma[T] = \sqrt{Var[T]} = \sqrt{\frac{\pi^2}{3} N^2 + O(N \ln N)} = O(N). 
\end{equation}

These expected values align with those of the Coupon Collector's Problem (CCP)\cite{MU2017}, which illustrates that the broadcast time mirrors the distribution of the number of coupons collected until all $N$ types of coupons are acquired in the CCP. 
Theorem 5.13 in \cite{MU2017} demonstrates that this distribution follows the Gumbel distribution when $N$ is sufficiently large (or mathematically $N \to \infty$).
\begin{equation}
        \lim_{N \to \infty} \textrm{Pr}( T > N \ln N + c N) = 1 - e^{-e^{-c}}, \label{eq:gumbel_ccdf}
\end{equation}
where $c$ is an arbitrary constant. 
The Gumbel distribution is known to be a probability distribution with a sharp peak near its mean value.

We herein provide a brief overview of the CCP. 
In this scenario, each purchase of a product containing a coupon results in one randomly selected coupon from $N$ types.
The objective is to determine how many purchases are required to obtain all $N$ types of coupons. 
Notably, the transition probability from state $i$ to $i+1$ in CCP is denoted by $p(i,i+1) = (N-i)/N$, in contrast to Eq.~(\ref{eq:transition_probability}) in the proposed stochastic model. 

Next, we briefly outline the derivation of the Gumbel distribution in CCP. 
Initially, the probability that the $i-$th coupon remains uncollected after $N \ln N + cN$ purchases is estimated as follows: 
\begin{equation}
        \left( 1 - 1/N \right)^{N(\ln N + c)} < e^{-(\ln N + c)} = e^{-c}/N.
\end{equation}
Given the independence of all coupon appearance events, the probability of acquiring any coupon can be calculated as follows when $N$ is sufficiently large.
\begin{equation}
        \left( 1 - e^{-c}/N \right)^{N} \simeq e^{-e^{-c}}.
\end{equation}
Taking the complementary event of the above event yields the Gumbel distribution as shown in Eq.~(\ref{eq:gumbel_ccdf}). 

Applying this result to the stochastic model of broadcast communications between validator nodes, with $E[T] = \mu = O(N \ln N)$, $\sigma[T] = \eta = O(N)$, and $(T > ) \tau_{broadcast} = \mu + c \eta = a N \ln N + c b N$ when the number of validator nodes $N$ is sufficiently large, we obtain the complementary cumulative distribution function of broadcast time as follows: 
\begin{equation}
        \bar{F}(\tau_{broadcast}) = 1 - e^{-e^{-(\tau_{broadcast} - \mu)/\eta}}, \label{eq:broadcast_ccdf}
\end{equation}
where $\mu= a N \ln N, \eta = b N$. 
By differentiating Eq.~(\ref{eq:broadcast_ccdf}) by $\tau_{broadcast}$, the probability density function of broadcast time is also obtained as follows: 
\begin{equation}
	f(\tau_{broadcast}) = \frac{1}{\eta} e^{-(\tau_{broadcast} - \mu)/\eta} e^{- e^{-(\tau_{broadcast} - \mu )/\eta}}. \label{eq:gumbel_distribution}
\end{equation}

\subsection{Distribution of block verification time}

Next, we examine the probability distribution of $\tau_{attest}$, representing the time taken to verify (or attest) the broadcast candidate block. 
Referring to Fig.~\ref{fig:bft_consensus}, we consider that the two phases, PREPARE (PRE-COMMIT) and COMMIT, correspond to the verification process. 
Since these phases can also be viewed as the time taken for the broadcast of block information, they can be assumed to follow the Gumbel distribution discussed in the previous section. 
We assume that both $\tau_{prepare}$ and $\tau_{commit}$ follow the same probability distribution as the Gumbel distribution, $f(\tau_{broadcast})$, previously discussed. 
Since $\tau_{attest} (= \tau_{prepare} + \tau_{commit})$ is the sum of these two phases, the probability distribution of $\tau_{attest}$ is derived by convolving two independent Gumbel distributions. 
Consequently, the following equation is obtained as the probability distribution of block verification time. 
\begin{eqnarray}
	& & f^{(2)}(\tau_{attest})  \nonumber \\
        &=& \int_{0}^{\tau_{attest}} f(\tau_{prepare}) f(\tau_{attest}-\tau_{prepare}) d\tau_{prepare} \nonumber \\
        &=& \int_{0}^{\tau_{attest}} \left( \frac{1}{\eta} e^{-(t-\mu)/\eta} e^{-e^{-(t-\mu)/\eta}} \right) \nonumber \\
        & & \times \left( \frac{1}{\eta} e^{-(\tau_{attest} - t -\mu)/\eta} e^{-e^{-(\tau_{attest} - t -\mu)/\eta}} \right) dt \nonumber \\
        &=& \frac{1}{\eta^2} e^{-(\tau_{attest}-2\mu)/\eta} \nonumber \\
        & & \times \int_{0}^{\tau_{attest}} e^{-( e^{^{-(t-\mu)/\eta}} + e^{^{-(\tau_{attest} - t -\mu)/\eta}} ) } dt, \nonumber \\
        & & \label{eq:attest_time_dist}
\end{eqnarray}
where $\mu = a N \ln N, \eta = bN$.

\subsection{Block time distribution}

Since the bottleneck determining block time is $\tau_{block} = \tau_{broadcast} + \tau_{attest} (>> \tau_{create})$, the probability distribution of block time can be expressed by convolving the three phases: PRE-PREPARE, PREPARE, and COMMIT. 
We assume that for each phase, it independently follows the Gumbel distribution.
Consequently, the following equation is obtained as the block time distribution. 
\begin{eqnarray}
	& & f^{(3)}(\tau_{block})  \nonumber \\
        &=& \int_{0}^{\tau_{block}} f(\tau_{block} - \tau_{attest}) f^{(2)}(\tau_{attest}) d\tau_{attest} \nonumber \\
        &=& \frac{1}{\eta^3} \int_{0}^{\tau_{block}} \left( e^{-(\tau_{block} - s - \mu)/\eta} e^{-e^{-(\tau_{block} - s - \mu)/\eta}} \right) \nonumber \\
        & & \times \left( e^{-(s-2\mu)/\eta} \int_{0}^{s} e^{-( e^{^{-(t-\mu)/\eta}} + e^{^{-(s - t -\mu)/\eta}} ) } dt \right) ds. \nonumber \\
        & & \label{eq:block_time_dist}
\end{eqnarray}

\subsection{Approximate representation of block time distribution}

Although Eq.~(\ref{eq:block_time_dist}) represents the block time distribution, its integral form can be inconvenient for theoretical analysis, particularly when considering fitting to real data. 
In this subsection, we aim to replace the integral term with an approximate calculation using the saddle--point method to derive an approximate representation of the block time distribution. 

Here, we consider a random variable representing the sum of $k$ random variables following the Gumbel distribution, denoted as $\tau^{(k)} = \sum_{l=1}^{k} T^{(gumbel)}_{l}$. 
Consequently, the probability density function $f^{(k)}(t; \mu, \eta)$ of this random variable can be approximated as follows: 
\begin{equation}
        f^{(k)}(t; \mu, \eta) \simeq \frac{t^{k-1}}{(k-1)! \eta^k} e^{-(t - k \mu)/\eta} e^{-k e^{-(t - k \mu)/(k \eta)}}, \label{eq:approximate_result}
\end{equation}
where $\mu = a N \ln N$, and $\eta = b N$. 
Equation (\ref{eq:approximate_result}) can be proven by combining the saddle--point method with induction. 
The detailed proof is provided in Appendix A. 

Using Eq.~(\ref{eq:approximate_result}), the probability distribution of the block verification time $\tau_{attest}$ is equivalent to $\tau^{(2)}$, which consists of two broadcast times, PREPARE (PRE-COMMIT) and COMMIT. 
Therefore, it can be approximated as follows.
\begin{equation}
        f^{(2)}(t; \mu, \eta) \simeq \frac{t}{\eta^2} e^{-(t - 2 \mu)/\eta} e^{-2 e^{-(t - 2 \mu)/(2 \eta)}}, \label{eq:approximate_result2}
\end{equation}
The probability distribution of block time $\tau_{block}$ is also equivalent to $\tau^{(3)}$, which consists of three broadcast times: PRE-PREPARE (PRE-VOTE), PREPARE (PRE-COMMIT), and COMMIT.
Therefore, it can also be approximated as follows: 
\begin{equation}
        f^{(3)}(t; \mu, \eta) \simeq \frac{t^{2}}{2 \eta^3} e^{-(t - 3 \mu)/\eta} e^{-3 e^{-(t - 3 \mu)/(3 \eta)}}, \label{eq:approximate_result3}
\end{equation}

\begin{figure}[t!]
  \centering
  \includegraphics[width=0.5\textwidth]{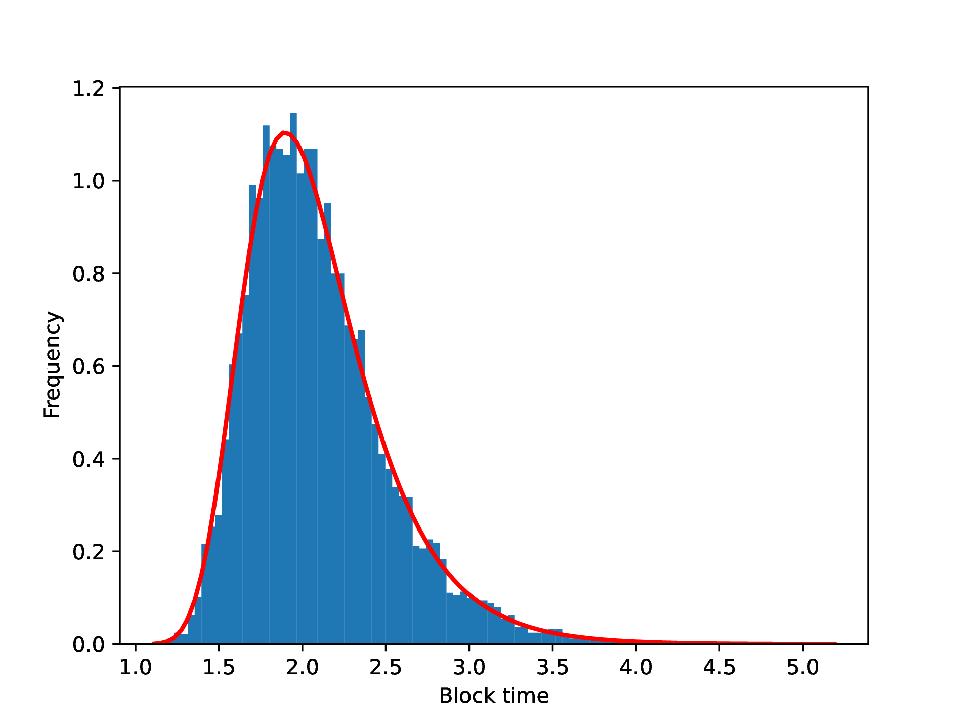}
  \includegraphics[width=0.5\textwidth]{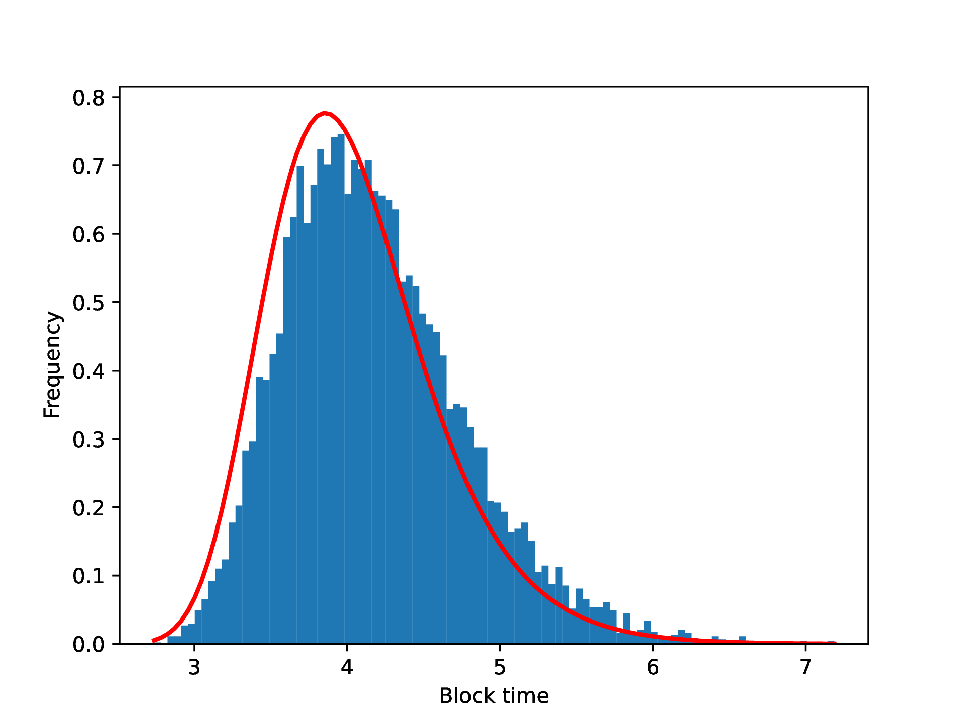}
  \includegraphics[width=0.5\textwidth]{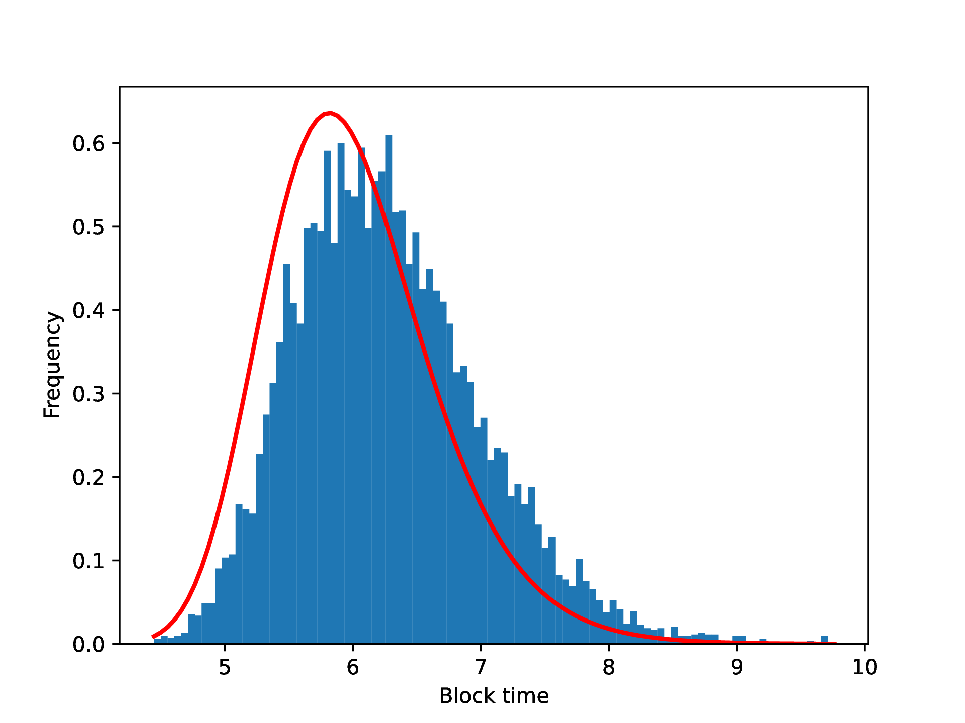}
	\caption{Results of numerical analyses to evaluate the theoretically derived formula $f^{(k)}$ for $k=1,2,3$. The histogram at the top illustrates a pattern of randomly sampled data according to the Gumbel distribution and the fitted curve of the approximate representation, $f^{(1)}$. One at the middle illustrates a pattern of randomly sampled data according to the sum of two independent Gumbel distributions and the fitted curve of $f^{(2)}$, and one at the bottom illustrates a pattern of randomly sampled data according to the sum of three independent Gumbel distributions and the fitted curve of $f^{(3)}$.}
  \label{fig:numerical_experiment}
\end{figure}

Next, we performed numerical analyses to evaluate the theoretically derived formula $f^{(k)}$ for $k=1,2,3$. 
The results are depicted in Fig. \ref{fig:numerical_experiment}; 
$f^{(1)}$ exactly matches the Gumbel distribution itself by definition. 
Since the approximate representations are calculated using the saddle--point method, we can observe that the red lines in Fig.~\ref{fig:numerical_experiment} are slightly shifted to the left for $k=2, 3$. 
Consequently, the approximate representations from theoretical analysis can reproduce the overall shape of the histogram from numerical analysis, and it is found to be effective approximations when $k$ is small. 
However, it is notable that the approximate representation tends to gradually shift to the left from the true probability distribution as $k$ increases. 
Therefore, caution must be exercised when using this approximate representation for large values of $k$.

\section{Discussion}\label{sec:discussion}

In this section, we present the deterministic expressions of coefficients $a, b$, which have not been detailed previously. 
Subsequently, we employ the result of theoretical analysis from Eq.~(\ref{eq:approximate_result3}) to fit the real data of Cosmos block time. 
Finally, the convolution of $k$ independent Gumbel distributions theoretically converges to the normal distribution when $k$ is sufficiently large, aligning with the central limit theorem\cite{DKLM2005}.

\subsection{Fitting the result of theoretical analysis with Cosmos block time data}

From Eqs.~(\ref{eq:T_average}) and (\ref{eq:T_variance}), we can derive the coefficients of the average, $\mu=a N \ln N$, and the standard deviation, $\eta=b N$, for the Gumbel distribution as follows. 
\begin{equation}
        a = 2 \Delta t, \hspace{5mm} b \simeq \frac{\pi}{\sqrt{3}} \Delta t, 
\end{equation}
where $\Delta t$ represents the average time taken to transfer one block to a randomly selected validator node. 

Substituting the number of validator nodes in Cosmos as $N=175$ \cite{IDA} into the theoretical result in Eq.~(\ref{eq:approximate_result3}), \textit{i.e.}, $f^{(3)}(t; \mu=a 175 \ln 175, \eta= b 175)$, we perform a non-linear fitting to the Cosmos block time data. 
We used a Python language module called \textbf{scipy.optimize.curve\_fit} \footnote{\url{https://docs.scipy.org/doc/scipy/reference/generated/scipy.optimize.curve_fit.html}} for the non-linear fitting. 
The result is depicted in Fig.~\ref{fig:data_blocktime}. 
As a result, we estimate $\mu \fallingdotseq 2.002896$ and $\eta \fallingdotseq 0.363636$. 
With $N=175$, a single block transfer time per step can be estimated as $\Delta t = \mu /(2N \ln N) = \eta / (\pi N / \sqrt{3}) \fallingdotseq 0.0011$s consistently. 
This outcome indicates that blocks are randomly transferred to other validator nodes at a frequency of approximately $1/\Delta t \fallingdotseq 909$ times per second on average. 
This outcome indicates that since $N$ validator nodes transfer blocks in parallel, each validator node transfers blocks randomly to other validator nodes at a frequency of approximately $1/(N \Delta t) \fallingdotseq 5.1948$ times per second on average. 
Verification of this communication frequency based on real data obtained from measurements using actual Cosmos nodes is awaited (See also Appendix B).

\subsection{Skewness of the convolution of many independent Gumbel distributions} 

It is known that the skewness of the Gumbel distribution is approximately 1.14 \cite{gumbel1954}. 
Through numerical analysis, we observe that the skewness of $f^{(k)}$ decreases monotonically with the increase of $k$. 
Additionally, we also found that the skewness of the distribution of Cosmos block time in Fig.~\ref{fig:data_blocktime} is approximately 0.79, indicating a deviation from that of the Gumbel distribution. 
As $k$ becomes sufficiently large, the skewness tends to approach zero, suggesting that it approaches a normal distribution with a skewness of zero. 
Therefore, it can be concluded that the convolution of many independent Gumbel distributions is generally different from the Gumbel distribution itself but approaches to the normal distribution, whose skewness is zero. 
Furthermore, from numerical analysis, we observe that the standard deviation of $f^{(k)}$ gradually increases with the increase of $k$. 

We can theoretically explain both the decrease in skewness and the increase in standard deviation using the approximate representation in Eq.~(\ref{eq:approximate_result}) as follows: 
\begin{eqnarray}
	& & \ln f^{(k)}(t; \mu, \eta) \nonumber \\
	&\simeq& (k-1) \ln t - \ln( (k-1)! ) -k \ln \eta - \frac{t-k \mu}{\eta} \nonumber \\
	& & - k e^{-(t-k \mu)/(k \eta)} \nonumber \\
	&\simeq& (k-1) \ln t - \frac{t}{\eta} -k \left\{ 1 - \frac{t-k\mu}{k\eta} + \frac{1}{2} \left( \frac{t-k\mu}{k\eta} \right)^2 \right. \nonumber \\
	& & \left. - \frac{1}{3!} \left( \frac{t - k \mu}{k \eta} \right)^3 + O(t^4) \right\} \nonumber \\
	&=& C + (k-1) \ln t  - \frac{1}{2k\eta^2} ( t-k\mu)^2 + \frac{1}{6 k^2} O(t^3), 
\end{eqnarray}
where $C=k(\mu/\eta -1)$.
It is evident from the above calculation that since $O(t^3)$ is related to the skewness, the skewness term decreases as $k$ increases. 
Furthermore, the following approximation is applicable in the region where $t$ is large.
\begin{equation}
	f^{(k)}(t; \mu, \eta) \simeq \exp\left\{- \frac{1}{2k\eta^2} ( t-k\mu)^2 \right\}.
\end{equation}
This outcome indicates that the block time distribution approaches the normal distribution when $k$ is sufficiently large. 
Moreover, the standard deviation of the normal distribution becomes $\eta\sqrt{k}$, suggesting that the standard deviation of $f^{(k)}$ increases monotonically with the increase of $k$.

A previous study\cite{EKJ2023} examined the distribution of the IBC message relay time, which involves transferring tokens between different blockchains in Cosmos. 
They elucidated that the IBC message relay time comprises three segments: (1) the time for propagating and committing the IBC message relay transaction, (2) the time between the IBC message relay transaction's commitment and the corresponding IBC receive packet transaction, and (3) the time between the IBC receive packet commitment and IBC acknowledgment transaction commitment. 
Each segment incorporates one block into blockchains, resulting in at least three block times during the IBC message relay time. 
Since the block time distribution in Byzantine fault--tolerant consensus blockchains is expressed by $f^{(3)}$, the time distribution of the three segments of IBC message relay might be associated with $f^{(3 \times 3)} = f^{(9)}$ at the minimum, meaning its fluctuation might be considerably large. 
Further research on this topic is eagerly anticipated.

\section{Conclusion}\label{sec:conclusion}

We proposed a stochastic model for block propagation and verification in Byzantine fault--tolerant consensus blockchains and presented the results of theoretical analysis on the block time distribution. 
Our model addresses block propagation and verification, demonstrating that the broadcast time among validator nodes generally follows the Gumbel distribution, as observed in the CCP. 
Subsequently, we derived the distributions of block verification time and block time in integral form by computing the convolution of the Gumbel distribution. 
However, because of the inconvenience of this integral form for theoretical analysis, we derived an approximate representation. 
Additionally, we validated our theory by fitting the theoretical results to real block time data, thereby estimating a consistent single block transfer time between validator nodes. 

For future work, it is imperative to validate the estimated results from data analysis in real--world scenarios. 
In the proposed model, we assumed that every validator node randomly and independently selected a validator node to transfer blocks. 
However, in practice, block transfer should be performed in a more efficient way, such as transferring blocks not randomly but sequentially to validator nodes that have not yet transferred. 
It is important to consider dynamic factors inherent in blockchain networks, such as changes in network latency and validator node availability, and varying transaction loads. 
Furthermore, it would be intriguing to explore the explanation of the IBC message relay time using our theoretical findings.

\appendices{\textbf{Appendix A: Proof of Eq.~(\ref{eq:approximate_result})}}
\label{appendix1}

Substituting $k=1$ into Eq.~(\ref{eq:approximate_result}), it is easily verified that it corresponds to the Gumbel distribution in Eq.~(\ref{eq:gumbel_distribution}), which confirms that Eq.~(\ref{eq:approximate_result}) is consistent with $k=1$. 
Subsequently, we calculate $f^{(k+1)}$ using the convolution of $f^{(k)}$ and $f^{(1)}$ under the condition that $f^{(k)}$ holds to follow the typical procedure for proof by induction.
\begin{eqnarray}
	& & \int_{0}^{t} f(t - t_1) f^{(k)}(t_1) dt_1 \nonumber \\
	&=& \int_{0}^{t} \left( \frac{1}{\eta} e^{-(t - t_1 - \mu)/\eta} e^{- e^{-(t - t_1 - \mu)/\eta}}  \right) \nonumber \\
	& & \times \left( \frac{t_1^{k-1}}{(k-1)! \eta^k} e^{-(t_1 - k \mu)/\eta} e^{- k e^{- (t_1 - k \mu)/(k \eta)}}  \right) dt_1 \nonumber \\
	&=& \frac{1}{(k-1)! \eta^{k+1}} e^{-(t-(k+1)\mu)/\eta} \nonumber \\
	& & \times \int_{0}^{t} t_1^{k-1} e^{-(k e^{-(t_1 - k \mu)/(k \eta)} + e^{-(t - t_1 - \mu)/\eta})} dt_1. \label{eq:app3_derive1}
\end{eqnarray}

Here, the saddle--point method is used, and the argument of the exponential function in the integral kernel is extracted as 
\begin{equation}
	Z(t_1; t) =  k e^{-(t_1 - k \mu)/(k \eta)} + e^{-(t - t_1 - \mu)/\eta}. \label{eq:app3_derive2}
\end{equation}
The integral kernel reaches its maximum at the extreme value with the lowest value of $t_1$. 
Therefore, we approximate it by replacing it with this extreme value. 
To achieve this, we differentiate $Z(t_1; t)$ with respect to $t_1$ and then substitute its extreme value into the integral kernel for integration. 
\begin{equation}
	\frac{d Z}{dt_1} = k \left( - \frac{1}{k \eta} \right) e^{-(t_1 - k \mu)/(k \eta)} - \left(- \frac{1}{\eta} \right) e^{-(t - t_1 - \mu)/\eta} = 0.
\end{equation}
We can solve this equation as $t_1 = kt/(k+1)$.
\begin{equation}
	t_1 = \frac{k}{k+1} t. 
\end{equation}
Substituting this into Eq.~(\ref{eq:app3_derive2}), we get
\begin{equation}
	Z(t_1=kt/(k+1); t) =  (k+1) e^{-(t - (k+1) \mu)/((k+1) \eta)}. \label{eq:app3_derive3}
\end{equation}
Substituting this into the argument of the exponential function in the integral in Eq.~(\ref{eq:app3_derive1}) yields: 
\begin{eqnarray}
	& & \int_{0}^{t} f(t - t_1) f^{(k)}(t_1) dt_1 \nonumber \\
	&=& \frac{ e^{-(t-(k+1)\mu)/\eta} }{(k-1)! \eta^{k+1}} \int_{0}^{t} t_1^{k-1} e^{- (k+1) e^{-(t - (k+1) \mu)/((k+1) \eta)}} dt_1 \nonumber \\
	&=& \frac{ e^{-(t-(k+1)\mu)/\eta} }{(k-1)! \eta^{k+1}} e^{- (k+1) e^{-(t - (k+1) \mu)/((k+1) \eta)}} \int_{0}^{t} t_1^{k-1} dt_1 \nonumber \\
	&=& \frac{t^k}{k! \eta^{k+1}} e^{-(t-(k+1)\mu)/\eta} e^{- (k+1) e^{-(t - (k+1) \mu)/((k+1) \eta)}} \nonumber \\
	&=& f^{(k+1)}(t; \mu, \eta). \label{eq:app3_derive4}
\end{eqnarray}
This calculation confirms that $f^{(k+1)}$ in Eq.~(\ref{eq:approximate_result}) can be derived successfully from the assumption of $f^{(k)}$ in Eq.~(\ref{eq:approximate_result}), which proves Eq.~(\ref{eq:approximate_result}) by induction.

\appendices{\textbf{Appendix B: Adjustment to take into consideration the number of validator nodes required for the consensus}}
\label{appendix2}

In PBFT--based distributed consensus algorithms, approval from $2f+1$ validator nodes out of $N=3f+1$ is sufficient to reach consensus and commit the finalized block.
Therefore, strictly, waiting for the block to be broadcast to all validator nodes may not be necessary. 
When considering this, the coefficients regarding the mean and standard deviation of the Gumbel distribution may change slightly. 

In the derivation of Eqs.~(\ref{eq:T_average}) and (\ref{eq:T_variance}), we considered the broadcast time, $T = \sum_{i=1}^{N-1} T_i$. 
Taking the above into account, the broadcast time can be adjusted to $\tilde{T} = \sum_{i=1}^{\lceil 2(N-1)/3 \rceil} T_i$. 
This adjustment does not affect the $N$ dependence; however, it replaces $N$ with $2N/3$ in $\mu=a N \ln N$ and $\eta=b N$ as follows: 
\begin{eqnarray}
	\mu  &=& \frac{2a}{3} N (\ln N + \ln(2/3) ) \simeq \frac{4 \tilde{\Delta t}}{3} N \ln N,  \\
	\eta &=& \frac{2b}{3} N = \frac{2\pi}{3\sqrt{3}} \tilde{\Delta t} N. 
\end{eqnarray}
By dividing $\Delta t \fallingdotseq 0.0011$s by a correction factor of $2/3$, single block transfer time between validator nodes becomes approximately $\tilde{\Delta t} \fallingdotseq 0.0011 / (2/3) = 0.00165$s. 
Consequently, the communication frequency of block transfer becomes $1/(N \tilde{\Delta t}) \fallingdotseq 5.1948 \times (2/3) \fallingdotseq 3.4632$ times per second. 

\end{document}